\documentclass{article}

\usepackage{amsmath,graphicx,hyphenat,natbib,rotating,fancyhdr}

\fancypagestyle{plain}{
  
  \fancyhead[L]{SYNTHESIS ARTICLE}
}

\begin{document}
\title{Generalized modeling of empirical social-ecological systems}
\author{Steven J. Lade$^{1,2}$ and Susa Niiranen$^1$}
\date{}
\maketitle

\noindent $^{1}$Stockholm Resilience Centre, Stockholm University, Kr\"aftriket 2B, 114 19 Stockholm, Sweden

\noindent $^{2}$Fenner School of Environment and Society, The Australian National University, Building 141,
Linnaeus Way, Canberra, ACT 2601, Australia

\vspace{8pt}
\noindent \emph{Note: This is the authors' accepted manuscript. The final version of the paper including copy-editing can be accessed at \\ http://onlinelibrary.wiley.com/doi/10.1111/nrm.12129/full}

\newpage
\section*{Abstract}
Modeling social-ecological systems is difficult due to the complexity of ecosystems and of individual and collective human behavior. Key components of the social-ecological system are often over-simplified or omitted. Generalized modeling is a dynamical systems approach that can overcome some of these challenges. It can rigorously analyze qualitative system dynamics such as regime shifts despite incomplete knowledge of the model's constituent processes. Here, we review generalized modeling and use a recent study on the Baltic Sea cod fishery's boom and collapse to demonstrate its application to modeling the dynamics of empirical social-ecological systems. These empirical applications demand new methods of analysis suited to larger, more complicated generalized models. Generalized modeling is a promising tool for rapidly developing mathematically rigorous, process-based understanding of a social-ecological system's dynamics despite limited knowledge of the system.

\subsection*{Recommendations for Resource Managers}
\begin{itemize}
\item Understanding empirical social-ecological dynamics requires integrating quantitative and qualitative data
\item Generalized modeling can analyze qualitative dynamics, such as regime shifts, by integrating both qualitative and quantitative data
\item Generalized modeling is well-suited to use in participatory or collaborative settings 
\end{itemize}

\vspace{1cm}

\noindent Keywords: generalized modeling; social-ecological systems; eigenvalues; dynamical systems

\newpage
\section{Introduction}
Humanity relies on the earth's biophysical processes for a range of `services' ranging from the fulfillment of basic physiological needs such as food and water to facilitating the cultural and spiritual interactions that help bind societies \citep{MEA__2005}. At the same time, these biophysical processes are being affected by humans on a dramatic scale \citep{Steffen2015,Steffen2015a}. To understand and manage biophysical systems such as natural resources, or to understand today's human welfare problems, therefore often requires acknowledging that these systems are part of interdependent social-ecological systems \citep{Berkes__1998}.

Understanding the dynamics and dynamical stability of social-ecological systems is especially important \citep{Folke_ES_2010}. For example, a sustainable social-ecological system may be associated with the stability (over time) of a desirable basin of attraction \citep{Derissen_EE_2011}. Dynamical stability has already been an object of research in theoretical ecology for decades, generally focusing on the relationship between stability and diversity or stability and food web structure \citep{Holling1973,may1972will,Scheffer2001,Ives_S_2007}. Stability is also closely connected to the recent research fields of regime shifts in ecosystems \citep{Scheffer2001,Scheffer2003}, in which stability is lost leading to large and abrupt changes in ecosystems, and early warning signals for critical transitions \citep{Dakos2012,Scheffer2009,Scheffer2003}, in which statistics of time series such as variance and autocorrelation are used to indirectly estimate stability \citep{Kuehn2011}. Dynamical stability is also receiving increasing attention in studies of human decision-making \citep{Kelso1997,Scherbaum2008}.

Dynamical systems models \citep{Kuznetsov_2010_src,Kelly2013} provide a convenient framework with which the dynamics and stability of systems can be assessed. Constructing suitable dynamical models of empirical social-ecological systems is a difficult undertaking, however \citep{Pahl-Wostl2007,Schlueter_NRM_2012}. Both individual and collective human behavior and ecological and biophysical dynamics are extremely complex and context-dependent and our knowledge of their interactions is highly partial and uncertain. Integrating human behavior with ecological dynamics is additionally complicated because the two domains often involve dramatically different theories and conceptual backgrounds as well as different types and amounts of available data \citep{Filatova2013}. Modeling approaches often deal with this complexity by over-simplifying the system, or by omitting one domain altogether \citep{Schlueter_NRM_2012}. The resulting models often are of only theoretical interest without clear validation from or significance for case studies, lack a complete understanding of processes driving the social-ecological system's dynamics, or stop at a qualitative level such as causal loop diagrams. 

Here, we show how the generalized modeling approach \citep{Gross2006,Kuehn2013} can deal with the often limited availability of data and knowledge on causal relationships in social-ecological systems, support integration of knowledge across disciplines and support the development of more realistic models while still permitting formal and rigorous analysis of the stability and qualitative dynamics of the social-ecological system. Primarily used in ecology \citep{Gross_Science_2009,Stiefs_TAN_2010,Yeakel_TE_2011,Yeakel2014}, generalized modeling is a type of dynamical systems approach that is well suited to research questions involving the roles of processes and feedbacks in dynamical patterns such as attractors, regime shifts, and oscillatory dynamics. A key advantage of a generalized modeling approach is that it does not require full specification of the functional forms of these processes. Social and ecological components can therefore be modeled and empirically grounded at comparable levels of complexity, despite differing levels of knowledge and data availability. We show that development of a generalized model is also well suited to a collaborative or participatory \citep{stave2010participatory} process, in which researchers or non-academic stakeholders with different areas of knowledge contribute to the modeling process. Applying generalized modeling to a complex social-ecological system, however, demands new methods of analyzing the generalized model, which we introduce in this article.

In a generalized modeling analysis, one takes the following steps \citep{Gross2006,Kuehn2013} (Fig. \ref{fig:GM}). (1) As in any modeling activity, a conceptual model of the system to be modeled should be identified. (2) A `generalized model' of the system should be constructed, in which state variables and processes affecting those state variables are written down in differential equation form, but functional forms of those processes are not identified. (3) The Jacobian matrix of the generalized model is symbolically calculated. (4) The entries in the Jacobian matrix are re-written in terms of the three `generalized parameters': the $\alpha$ parameters, $\beta$ parameters, and elasticities. (5) Values or ranges are assigned to these parameters. (6) As in any modeling exercise, the model ought to be validated in some way and its results can then be analyzed.

Analyzing a dynamical system by studying qualitative changes in its behavior has a long history across multiple disciplines \citep{puccia2013qualitative,Kuznetsov_2010_src,Kelso1997,strogatz2014nonlinear,andronov1973theory,guckenheimer2013nonlinear,Scheffer2001,Scherbaum2008}. The use of Jacobian matrices or similar objects as a specific method to study stability near fixed points is also well established \citep{novak2016characterizing,bender1984perturbation,Kuznetsov_2010_src,strogatz2014nonlinear,guckenheimer2013nonlinear}. Estimating the entries in the Jacobian matrix, which is essentially a matrix of partial derivatives, can however be difficult in settings where the exact functional form of the dynamical systems model is not known. The novelty of the generalized modeling approach is steps (2) and (4) of the procedure outlined above: the use of the `generalized parameters', which are easily and intuitively interpreted, to directly parameterize the Jacobian matrix; and the structuring of the entries of the Jacobian matrix according to the generalized model. In this paper, we make the new claim that generalized modeling is particularly useful for understanding empirical social-ecological systems.

We illustrate the application of the generalized modeling approach using recent work we undertook on modeling the collapse of the cod fishery in the Baltic Sea \citep{LadeRESULTS}. In the late 1970s to early 1980s, the biomass of cod in the Baltic Sea reached previously unseen high levels. The cod stock persisted at this high level for several years before suddenly and dramatically collapsing in the late 1980s to levels from which there has been only partial recovery \citep{Eero2008,Mollmann2009,SOU1993}. During the boom and collapse, regulation of the fishery was ineffective or non-existent, but government subsidies encouraged expansion of the fishing fleet \citep{SOU1993}. In the study, we sought to understand the role of social processes in the boom and collapse of the social-ecological system constituted by the cod fishery, in contrast to the usually exclusively ecological studies of the system. Specifically, we focused on factors affecting how fishers make decisions, such as: sunk cost effects, the tendency to
continue an endeavor once an investment in money, effort, or
time has been made \citep{Arkes1985,Janssen2004}; delays in decision-making and in update of perceptions of the fishery's condition; fishing in the Baltic of an external fleet; and the effects of government subsidies and market price. Generalized modeling is well suited to such research questions, because of its focus on processes and feedbacks and because persistence and risk of collapse can be readily interpreted in terms of stability.

Apart from the Baltic Sea study \citep{LadeRESULTS}, we have also used generalized modeling to aid the construction of an early warning signal \citep{Lade2012} and as a tool to demonstrate that social-ecological interactions can generate regime shifts in social-ecological systems \citep{Lade2013}. In the early warning signal study, the constraints on the Jacobian matrix provided qualitative knowledge about the system, encoded through into the Jacobian by the generalized modeling approach, were used to reduce the amount of data required to calculate the early warning signal. In the social-ecological regime shift study, generalized modeling was used as an approach to study the bifurcations of a entire class of dynamical system models, that is, all those models that had the same generalized model structure.

We first describe how to construct a generalized model of an empirical social-ecological system, including the advantages that generalized modeling confers over conventional dynamical models. Using the Baltic cod fishery case as an example, we describe the particular advantages and challenges involved in applying generalized modeling to empirical social-ecological systems. We then introduce several methods that can be used to analyse a generalized model of an empirical social-ecological system. Some of these methods are described here for the first time in detail. Finally, we reflect broadly on our experience using generalized modeling for the Baltic cod fishery case study and the potential of generalized modeling for empirical social-ecological systems more broadly.

\section{Model construction}
\begin{figure}
\begin{center}
\includegraphics{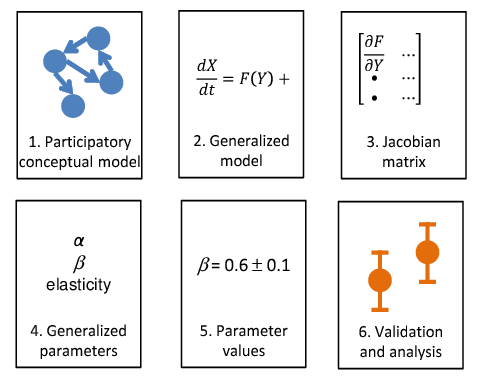}
\end{center}
\caption{Summary of the generalized modeling procedure.\label{fig:GM}}
\end{figure}

In this section, we outline the steps involved in the construction and parameterization of a generalized model of an empirical social-ecological system (Fig \ref{fig:GM}). Our description is mostly qualitative; the formal mathematical foundation of generalized modeling has been described many times elsewhere \citep{Gross2006,Kuehn2013, Lade2013}.

\subsection{Conceptual model}
The first task in a generalized modeling analysis is similar to that for any dynamical model. The structure of the model to be constructed needs to be identified: the important state variables in the system, the processes which affect those state variables, and with which state variables these processes interact. In previous literature on generalized modeling this step has generally been overlooked, and an appropriate model simply assumed. Modeling social-ecological systems, however, requires a large variety of disciplinary knowledge, and modeling an empirical case requires careful decisions about research questions, system boundaries, and the level of detail of the description.

A useful tool for model conceptualization from system dynamics is the causal loop diagram \citep{Sterman2000}. Causal loop diagrams conceptualize systems in terms of stocks or variables and interactions between those variables, and therefore serve as a good basis from which to identify the state variables and processes needed for a generalized model. Causal loop diagrams are already widely used to assess feedbacks within social-ecological systems and many researchers are therefore already familiar with them.

\begin{figure*}
\begin{center}
\includegraphics{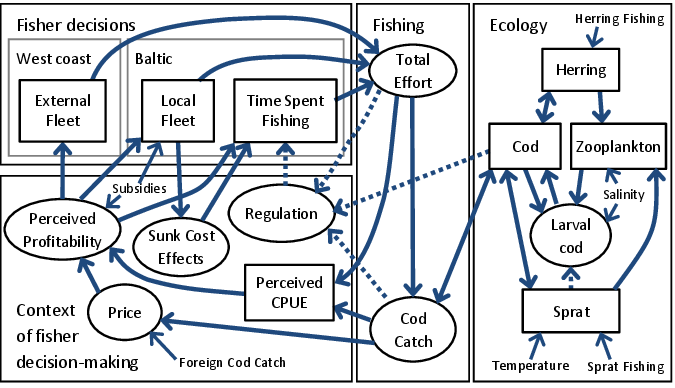}
\end{center}
\caption{Simplified version of the collaboratively developed causal loop diagram for the Baltic cod fishery social-ecological model. Arrows indicate the direction of influence of one quantity on another. Boxes indicate state variables treated as stocks; ovals denote intermediate variables; smaller text indicates drivers. Dotted lines indicate interactions included in hypothetical model `experiments'. CPUE denotes Catch Per Unit Effort. Modified version of \citet{LadeRESULTS}.\label{fig:Baltic}}
\end{figure*}

In the Baltic study, a group with expertise on the ecological, political, economic, and sociological aspects of the Baltic Sea was assembled. The group developed a collective causal loop diagram outlining the important entities and interactions affecting the dynamics of the Baltic cod fishery during the 1980s \citep{LadeRESULTS} (simplified version shown in Fig. \ref{fig:Baltic}).

To illustrate the model construction process, let us focus on the processes driving a small part of the generalized model of the Baltic Sea social-ecological system: the state variable CPUEPerception (Fig \ref{fig:Baltic}) representing the fishers' perception of the state of the fishery. We assume that this perception lags behind the actual state of the fishery, for example due to social identity \citep{Sonvisen2014} or risk aversion \citep{dowling2015risk}.

\subsection{Generalized model}
The state variables and processes graphically represented in the conceptual model need to be mathematically represented in a generalized model. In this context, a generalized model is a system of differential equations in which functional forms are not specified, but represented by `placeholder' functions. For example, in the Baltic cod fishery study, the equation for the dynamics of the CPUEPerception state variable described above had the following form (the complete generalized model is described in \citet{LadeRESULTS}):
\begin{equation}
\label{eq:CPUE1}
\frac{d\text{CPUEPerception}}{dt} = \text{CPUERelaxation}\left(\text{CPUE} - \text{CPUEPerception}\right).
\end{equation}
This equation implements the lagged nature of CPUEPerception (fishers' perception of CPUE) with respect to CPUE (Catch Per Unit Effort). The variable CPUEPerception will relax towards the actual CPUE at a rate determined by the function CPUERelaxation.

In a conventional dynamical systems model, each process is given a specific functional form (a Holling type-II function, for example). All the parameters of each functional form must also be specified. The dynamical system is then usually numerically simulated to find a fixed point of the system, and the system's stability calculated there. In generalized modelling, the Jacobian matrix of the system at a fixed point is instead directly parameterized, as described in the next section. This confers two advantages to a generalized model over a conventional dynamical systems model: first, less data is required to parameterize the model; second, less computation time is required to calculate stability. These advantages allow a generalized modeling approach to deal with the limited availability of data and knowledge that often occurs for a social-ecological system. On the other hand, because functional forms are not specified, a generalized model cannot be used to produce time series like a conventional dynamical systems model. 

\subsection{Jacobian matrix}
\label{sec:JM}
The next step in the generalized modeling procedure is a formal mathematical one. The existence of a fixed point \citep{Kuehn2013} is assumed and the Jacobian matrix of the generalized model is symbolically calculated at the fixed point.

Formally, all state variables are rescaled in terms of their values at the fixed point \citep{Kuehn2013}. (If $X$ has values $X^*$ at the fixed point, the change of variables $X \to X/X^*$ is performed.) This step does not affect the values of the generalized parameters introduced below, nor as was recently recognized \citep{Lade2013} is it needed in order to calculate eigenvalues. Rescaling is advantageous because it renders the values of state variables at the fixed points unnecessary for calculating eigenvectors and the other generalized modeling outputs described below. Elements of the Jacobian's eigenvectors, after rescaling, become expressed in fractional (fraction of the value of the state variable at its steady state) rather than absolute units.

To calculate the Jacobian matrix, we need the derivatives of the right-hand-side of the dynamical system's equations with respect to each state variable. Let us consider Eq.~\eqref{eq:CPUE1}. We begin by taking the total differential of the right-hand-side.
\begin{align}
d\text{RHS}_\text{CPUEPerception} &\equiv d\text{CPUERelaxation}\left(\text{CPUE} - \text{CPUEPerception}\right) \nonumber \\
&= \left.\frac{d\text{CPUERelaxation(x)}}{dx}\right|_{x = \text{CPUE} - \text{CPUEPerception}} \\ &\times \left(d\text{CPUE} - d\text{CPUEPerception}\right). \label{eq:CPUE2}
\end{align}
CPUE is, however, only an intermediate variable, itself depending on other state variables. To identify all the entries in the Jacobian, we need to continue until we have only differentials of state variables. The next step is to use the definition of CPUE,
\begin{equation}
\text{CPUE} =  \frac{\text{CodCatch}}{\text{FishingEffort}},
\end{equation}
and take total differentials:
\begin{equation}
\label{eq:dCPUE}
d\text{CPUE} = \text{CPUE} \times \left(\frac{d\text{CodCatch}}{\text{CodCatch}} - \frac{d\text{FishingEffort}}{\text{FishingEffort}}\right),
\end{equation}
and substitute into Eq.~\eqref{eq:CPUE2}. For illustrative purposes, let us take one more step. The definition of CodCatch is
\begin{multline}
\text{CodCatch} = \text{AdultCodCatch}(\text{FishingEffort}, \text{AdultCodBiomass}) \\
+ \text{JuvenileCodCatch}(\text{FishingEffort}, \text{JuvenileCodBiomass}).
\end{multline}
Its total differential is
\begin{multline}
\label{eq:dCCB}
d\text{CodCatch} = \frac{\partial\text{AdultCodCatch}}{\partial\text{AdultCodBiomass}}d\text{AdultCodBiomass} \\
+ \frac{\partial\text{JuvenileCodCatch}}{\partial\text{JuvenileCodBiomass}} d\text{JuvenileCodBiomass}\\
+ \left(\frac{\partial\text{AdultCodCatch}}{\partial\text{FishingEffort}} + \frac{\partial\text{JuvenileCodCatch}}{\partial\text{FishingEffort}}\right)d\text{FishingEffort}
\end{multline}
This process should be repeated until only differentials of state variables appear on the right-hand side, from which the derivatives needed for the Jacobian can be read off. In the Baltic Sea model, this process was partially automated.

Although a fixed point is assumed, the social-ecological system need not however reside precisely at the fixed point. The social-ecological system may fluctuate within a basin of attraction about the fixed point; the stability of the fixed point is used as a measure of the stability of the basin of attraction \citep{Scheffer2003}. The value of the fixed point may also slowly change in time due to the effects of what are called slow variables and drivers in the resilience literature \citep{Walker2012}, and the fast-slow time scale decomposition in mathematics \citep{Kuehn2011}. In the case of the Baltic fishery, a fixed point assumption can be justified by examining annual fish stocks (and fishermen behaviour) and thereby neglecting the seasonal variations that can lead to periodic behavior.

\subsection{Generalized parameters}
The entries in the Jacobian are next re-written in terms of the more easily interpreted `generalized parameters'. Here, we use three types of generalized parameters:
\begin{itemize}
\item $\alpha$ parameters, which represent the characteristic time scale for changes in a state variable. 
\item $\beta$ parameters, which compare the strengths of multiple processes that are acting on a state variable. Usually, all the processes with positive sign are compared against each other and all the processes with negative sign are compared.
\item Elasticity parameters, which characterize the non-linearity of the process. They take the general form
\begin{equation}
\label{eq:elasdef}
H_X = \frac{X}{H}\frac{\partial H}{\partial X}
\end{equation}
for the function $H(X)$ with respect to state variable $X$. For example, a function $H(X) = aX^n$ has elasticity $n$ for all values of $x$, while the Holling type II function $H(X) = aX/(b + X)$ has elasticity in the range $(0,1)$ for $b,X > 0$.
\end{itemize}

For Eqs.~(\ref{eq:CPUE2}-\ref{eq:dCCB}), a suitable choice of generalized parameters is
\begin{align}
\label{eq:alpha} \alpha[\text{CPUEPerception}] &= \left.\frac{\text{CPUERelaxation}(x)}
{x}\right|_{x = \text{CPUE} - \text{CPUEPerception}} \\
\label{eq:beta} \beta[\text{CodCatch}] &= \frac{\text{AdultCodCatch}}{\text{AdultCodCatch} + \text{JuvenileCodCatch}}
\end{align}
\begin{multline}
\label{eq:elas} \text{Elas[CPUERelaxation]} =  \frac{d\text{CPUERelaxation(x)}}{dx} \times \\ \left.\frac{x}{\text{CPUERelaxation(x)}}\right|_{x = \text{CPUE} - \text{CPUEPerception}}.
\end{multline}
\begin{align}
\label{eq:elas2} \text{Elas[JCC\textsubscript{FE}]} &=  \frac{\partial\text{JuvenileCodCatch}}{\partial\text{FishingEffort}} \frac{\text{FishingEffort}}{\text{JuvenileCodCatch}}. \\
 \label{eq:elas3} \text{Elas[ACC\textsubscript{FE}]} &=  \frac{\partial\text{AdultCodCatch}}{\partial\text{FishingEffort}} \frac{\text{FishingEffort}}{\text{AdultCodCatch}}. \\
  \label{eq:elas4} \text{Elas[JCC\textsubscript{JCB}]} &=  \frac{\partial\text{JuvenileCodCatch}}{\partial\text{JuvenileCodBiomass}} \frac{\text{JuvenileCodBiomass}}{\text{JuvenileCodCatch}}. \\
   \label{eq:elas5} \text{Elas[ACC\textsubscript{ACB}]} &=  \frac{\partial\text{AdultCodCatch}}{\partial\text{AdultCodBiomass}} \frac{\text{AdultCodBiomass}}{\text{AdultCodCatch}}.
\end{align}
The parameters can be readily interpreted as follows. $\alpha[\text{CPUEPerception}]$ is the rate at which perceptions of CPUE are updated (in units of 1/years). $\beta[\text{CodCatch}]$ is the proportion of the cod catch biomass contributed by adult fish (as opposed to juveniles). \text{Elas[CPUERelaxation]} denotes the elasticity of CPUERelaxation with respect to its argument, at the current operating point. The following four elasticities (Eq. \ref{eq:elas2}-\ref{eq:elas5}) specify the sensitivity of adult and juvenile cod catch to fishing effort and to the respective cod stocks. Using the generalized parameter definitions in Eqns. (\ref{eq:alpha}-\ref{eq:elas}) and the differentials in Eqs.~(\ref{eq:dCPUE},\ref{eq:dCCB}) the differential $d\text{RHS}_\text{CPUEPerception}$ in Eq.~\eqref{eq:CPUE2} can be rewritten as:
\begin{multline}
d\text{RHS}_\text{CPUEPerception} = \alpha[\text{CPUEPerception}] \times \text{Elas[CPUERelaxation]} \\ \times  \left(\text{CPUE} \left(\beta \text{Elas[ACC\textsubscript{FE}]} + (1-\beta) \text{Elasticity[JCC\textsubscript{FE}]} - 1\right)\frac{d\text{FishingEffort}}{\text{FishingEffort}} \right. \\ 
+ (1-\beta) \times \text{CPUE} \times \text{Elas[JCC\textsubscript{JCB}]} \frac{d\text{JuvenileCodBiomass}}{\text{JuvenileCodBiomass}} \\
\left. + \beta \times \text{CPUE} \times \text{Elas[ACC\textsubscript{ACB}]} \frac{d\text{AdultCodBiomass}}{\text{AdultCodBiomass}} - d\text{CPUEPerception}\right).
\end{multline}
We note that this expression contains no explicit derivatives of functions. Only the $\alpha$, $\beta$ and elasticity generalized parameters (and, once the derivation is complete, state variables that have been normalized to unity) need to be estimated.

The generalized parameters are parameters like in a conventional simulation model, but they parameterize the Jacobian matrix directly. Consequently, the generalized parameters can parameterize the entire class of models that give rise to that Jacobian matrix, rather than just a single model formulation as is usually the case in conventional models \citep{Kuehn2011}. Therefore a generalized modeling approach can effectively `scan' an entire set of dynamical system models that are consistent with an empirical social-ecological system, rather than formulating and choosing a specific model.

Because the generalized parameters are easily interpreted, they are also well-suited to integrating quantitative or even qualitative knowledge from various disciplines. $\beta$ parameters, for example, are a simple measure of the relative influence of two processes on another variable. Equal influence of the two processes corresponds to a $\beta$ value of 0.5 (for example, if AdultCodCatch = JuvenileCodCatch in Eq.~\eqref{eq:beta}).

\subsection{Parameterization}
\label{sec:parameterization}
Values or ranges are then estimated for the generalized parameters. In the Baltic cod fishery study, a range of social (such as fleet composition, subsidy and income data) and ecological (such as catch and diet data) data were used to estimate generalized parameters \citep{LadeRESULTS}. Generalized parameters were estimated for two periods: during the cod boom, and at the start of the cod collapse. We were careful not to apply the generalized model to other periods of the cod collapse, as the fishery was not at or near a stable state during this time.

In the Baltic cod fishery study, the available data for the generalized parameters fell into four broad categories. In each case, a value was assigned to the generalized parameter together with a range specifying a uniform distribution that was later used for uncertainty analysis.
\begin{itemize}
\item Parameters for which annual data was available, mostly ecological parameters based on catch and stock assessment data. Estimates of the generalized parameter were calculated from each year. The generalized parameter was assigned the mean of the estimates with a range given by the extremes of the estimates.
\item Parameters for which only qualitative information or a single quantitative estimate, such as fleet composition from a single year, was available. Parameter values and distributions were assigned on a case by case basis. For example, sometimes although quantitative data was available it was known to be an underestimate. The parameter's uncertainty range was therefore extended upwards accordingly.
\item Some parameters were given by the definition of variables in the model. For example, total effort is defined by number of vessels multiplied by time spent fishing
\begin{equation}
\text{TotalEffort} = \text{TotalVessels} \times \text{TimeSpentFishing}.
\end{equation}
The elasticities of total effort with respect to number of vessels and time spent fishing are therefore both unity, with zero uncertainty range:
\begin{align}
\text{TotalEffort}_\text{TotalVessels} &= \frac{\text{TotalVessels}}{\text{TotalEffort}} \frac{d\text{TotalEffort}}{d\text{TotalVessels}} = 1\\
\text{TotalEffort}_\text{TimeSpentFishing} &= \frac{\text{TimeSpentFishing}}{\text{TotalEffort}} \frac{d\text{TotalEffort}}{d\text{TimeSpentFishing}} = 1,
\end{align}
where we have used the definition of elasticity in Eq. \eqref{eq:elasdef}.
\item For the small number of parameters for which no empirical or theoretical data was available, a uniform distribution was assigned. For $\beta$ parameters, which can vary between 0 and 1, we used a uniform distribution over the full range 0 to 1. For elasticities, the value 1 represents a default assumption in which the relationship is linear, we chose a uniform distribution over 0.5 to 2 surrounding this default value.
\end{itemize}

\section{Model validation and analysis for large generalized models}
\label{sec:analysis}
We now describe methods of validation and analysis suitable for a large, empirically based generalized model such as the Baltic Sea cod fishery case (step 6 of Fig. \ref{fig:GM}). The methods include both long-standing approaches and methods first used on the generalized model of the Baltic Sea cod fishery \citep{LadeRESULTS}. In addition to estimating stability via eigenvalues (Sec.~\ref{sec:eigenvalues}), we introduce analyses that explore the contributions to social-ecological system stability of subsystems (Sec.~\ref{sec:decomposition}), feedbacks (Sec.~\ref{sec:feedback}), and individual links or parameters (Sec.~\ref{sec:sensitivity}). Eigenvector analysis allows a shift of focus to the roles of individual state variables (Sec.~\ref{sec:eigenvector}); uncertainty analysis quantifies the degree of confidence in model outputs (Sec.~\ref{sec:uncertainty}); and model experiments can be used to explore the consequences hypothetical modeling assumptions (Sec.~\ref{sec:experiments}).

Results from these analyses can also be used to validate the generalized model, by comparing the model's predicted patterns against the observed dynamics. Because a generalized model does not produce time series output like a conventional simulation model, validation of the model must proceed differently. The output of a generalized model instead consists of predictions of dynamical patterns: stable fixed points, loss of stability, oscillatory dynamics, and so on. (In this sense, generalized modeling is somewhat similar to the pattern-oriented modeling of agent-based models \citep{Grimm1996,Grimm2005}.) Stability of fixed points, as measured via eigenvalues, were the primary validation tool for the Baltic cod fishery model, as described in the next section. Eigenvectors were also used for validation, while all other model outputs were used for analysis.

\subsection{Eigenvalues}
\label{sec:eigenvalues}
A key output of a generalized modeling analysis is the eigenvalues of the Jacobian matrix. Eigenvalues are commonly used, both in conventional dynamical models and generalized models, to indicate stability of fixed points \citep{Kuznetsov_2010_src}. Here, we describe how eigenvalues of the Jacobian matrix can be interpreted for a generalized model of a social-ecological system, and how in the Baltic Sea case study eigenvalues were also used to validate the generalized model.

\subsubsection{Theory}
There are many possible definitions of stability \citep{Grimm1997,Ives_S_2007}. In this article, we use the asymptotic definition of stability, where (roughly speaking) a fixed point is stable if it returns to the fixed point after a small perturbation \citep{Kuznetsov_2010_src}. For a continuous time model formulation, as was used in the Baltic cod fishery model, if the real parts of all eigenvalues at a fixed point are negative then the fixed point is stable. A purely real dominant eigenvalue crossing the imaginary axis (that is, acquiring positive real part) corresponds to a fold bifurcation in the attractor landscape of the system, which is in turn often associated with regime shifts such as fishery collapses\footnote{A pair of complex conjugate eigenvalues crossing the imaginary axis corresponds to a Hopf bifurcation \citep{Kuznetsov_2010_src}, which is often associated with the emergence of oscillations such as in a predator-prey system \citep{Fussmann_S_2000}. It is a fold bifurcation, however, that we found in the Baltic cod fishery case.} \citep{Biggs_2012,Scheffer2003}. The eigenvalue with largest real part is therefore an indicator of stability (more precisely, instability, since a positive real part corresponds to an unstable fixed point).

Previous analyses of large generalized models \citep{Gross_Science_2009,Zumsande_JTB_2010,Yeakel2014} used a dichotomous measure of stability: Those combinations of parameters in which the dominant eigenvalue had positive real part were labeled `unstable', otherwise the parameter combination was `stable'. The tendency of systems to be stable across a range of generalized parameters was studied. In cases where the modeling aim is instead to estimate stability in a specific empirical case, however, such a dichotomous stability measure may be unable to identify changes in the stability of the system with sufficient detail. Consider estimating the eigenvalues of a system at several moments in time before a fold bifurcation. With a dichotomous measure, the decline in stability could not be measured; only the final collapse would be reported. Indeed, even the eigenvalues estimated from simulations of a theoretical system undergoing a fold bifurcation may never actually display positive real part because the fast and slow time scales are not sufficiently well separated \citep{Lade2012}. For empirical application to specific social-ecological systems, such as in the Baltic cod fishery case, we recommend using the real part of the dominant eigenvalue as an indicator of stability.

A continuous stability measure is, however, susceptible to so-called `localized modes' of the social-ecological system, which a dichotomous measure is not. Localized modes are dynamics involving a small part of the full social-ecological system that are often a result of symmetries in the model's construction \citep{Aufderheide2012}. They are often not representative of the dynamics of the full social-ecological system, but nevertheless can lead to an eigenvalue at or near zero that registers as the dominant eigenvalue. Systematically detecting and assessing localized modes, and selecting appropriate eigenvalues as stability indicators, is ongoing work \citep{Aufderheide2012} that is important for the future development of the generalized modeling approach. Alternatively, localized modes with zero eigenvalues may indicate that inappropriate modeling assumptions are being made \citep{Gross_JoTB_2009}.

Finally, the use of eigenvalues is frequently motivated by noting that regime shifts are often associated with fold bifurcations, as discussed above. There are according to the resilience literature two recognized mechanisms, however, by which a system can `tip' or undergo a `regime shift' into a new state \citep{Biggs_2012}. In one mechanism, drivers change the stability landscape of the system through a bifurcation so that the previously stable state disappears and the system must transition into a new state. In a second mechanism, not necessarily associated with a bifurcation, drivers cause a rapid shock directly to state variables of a system that push it out of a stable state into a new state. Following \citet{Ashwin2012}, we refer to these two mechanisms as B-tipping (for bifurcation) and N-tipping (for noise) respectively\footnote{A third mechanism of tipping, where drivers cause a rapid shift in the position of the basin of attraction \citep{Scheffer2001}, can also be associated with increased eigenvalue \citep{Kefi2013} and therefore we expect it to be detected by the methods for B-tipping.}. N- and B-tipping are of course not entirely independent. A system approaching a B-tip is also at increased risk of an N-tip, and is likely to tip due to noise before the bifurcation point is actually reached. However these two theoretical tipping mechanisms focus attention on different criteria and methods for validation and analysis of the model, as described here and also in relation to the eigenvector tests below.

\subsubsection{Use of eigenvalues for model validation}
\begin{figure}
\begin{center}
\includegraphics{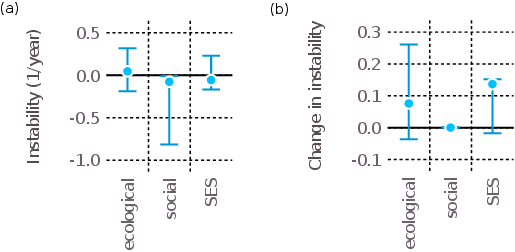}
\end{center}
\caption{\textbf{Eigenvalues and system decomposition for the generalized model of the Baltic Sea cod fishery.} (a) Dominant eigenvalues (instabilities) at the cod boom and (b) changes in dominant eigenvalues at the start of the collapse compared to the cod boom in the ecological, social and coupled social-ecological systems. Reproduced from \citet{LadeRESULTS}.\label{fig:results}}
\end{figure}

A generalized model cannot produce the time series output that is commonly associated with dynamical system models, which is commonly used for validation of the model. Instability, as measured by eigenvalues, was instead used as a validation test for the model of the Baltic cod fishery \citep{LadeRESULTS}. The fishery model was expected to have been stable during the cod boom, since the boom persisted as long as the longest time scale in the model, the reproductive cycle of Baltic cod. Stability was indeed predicted by the model (Fig \ref{fig:results}a, SES column). Another validation test was the change in stability of the fishery at the start of the collapse, at which point the instability should have increased. The fishery model indeed showed increased instability, in fact passing the threshold from stable to unstable, at the beginning of the collapse (Fig \ref{fig:results}b, SES column). This was strong support for the validity of the model, and for a regime shift via B-tipping in particular.

\begin{figure}
\begin{center}
\includegraphics{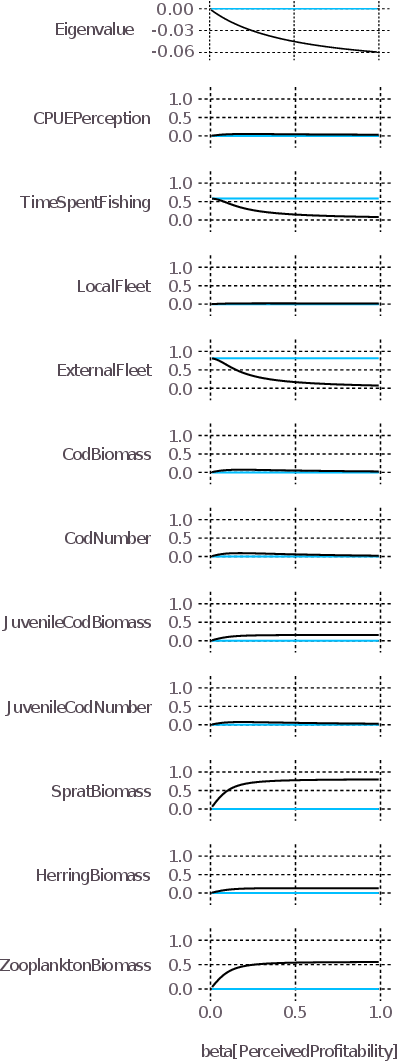}
\end{center}
\caption{\textbf{Localized eigenvectors.} (top row) Two eigenvalues of the Baltic Sea cod fishery generalized model \citep{LadeRESULTS} and (other rows) the absolute values of the components of their eigenvectors, as a function of the generalized parameter beta[PerceivedProfitability]. One of the eigenvectors (blue line) is a localized eigenvector (localized to the state variables TimeSpentFishing and ExternalFleet) for all values of the parameter. For intermediate values of the parameter, the other eigenvector (black line) is delocalized across both the ecological and social parts of the system. At high or low values of the parameter, however, the eigenvector becomes localized to two state variables in the ecological or social parts of the system, respectively.\label{fig:structuralzero}}
\end{figure}

As anticipated above, the Baltic cod fishery model indeed displayed a localized mode, due to symmetry between two state variables, as well as a second mode that became localized during model experiments (Fig \ref{fig:structuralzero}). In the Baltic cod fishery model the task of selecting the appropriate eigenvalue to use for reporting stability was however straightforward. After eliminating the zero eigenvalue of the localized mode, the next dominant eigenvalue was a purely real eigenvalue that became unstable at the start of the collapse \citep{LadeRESULTS}.

\subsection{System decomposition}
\label{sec:decomposition}
Social-ecological systems, as a framework for analysis, are believed to be important because of the feedbacks and the dynamics that emerge due to the coupling between the social and ecological systems \citep{Levin_EDE_2012,Berkes__1998}. Generalized modeling allows the importance of the social-ecological coupling to be simply tested by decomposing the system into decoupled social and ecological systems and analyzing the behavior of these decoupled systems.

Formally, submatrices (subsets of the full matrix's rows and columns) corresponding to the social and ecological subsystems are extracted from the full social-ecological system's Jacobian matrix. The submatrices obtained correspond to the Jacobian matrices of subsystems with fixed input from the rest of the system. Exploring an alternative fixed point, such as complete disconnection with zero input from the other subsystem, would likely involve Jacobian matrices with different values\footnotemark. The submatrices obtained can then be subjected to any of the analyses described in this section.

In the Baltic cod fishery, the submatrices' eigenvalues were calculated to assess the contributions of the decoupled systems to the social-ecological system's overall stability. The decoupled ecological (respectively, social) systems correspond to that subsystem with constant fishing effort from the social system (respectively, constant cod availability from the ecological system). Results demonstrated that during the boom the decoupled ecological system may in fact have been unstable, only yielding a stable coupled social-ecological system due to the stabilizing effect of the social system (Fig \ref{fig:results}a). Results also showed that the instability in the coupled social-ecological system may, remarkably, have increased more than either of its component sub-systems (Fig \ref{fig:results}a). These results are discussed at length in \citet{LadeRESULTS}.

\footnotetext{Formally, for a system comprised of two subsystems $\mathbf{x}$ and $\mathbf{y}$ such that $d\left[ \begin{smallmatrix} \mathbf{x} \\ \mathbf{y} \end{smallmatrix} \right] /dt = \left[ \begin{smallmatrix} \mathbf{f}(\mathbf{x},\mathbf{y}) \\ \mathbf{g}(\mathbf{x},\mathbf{y}) \end{smallmatrix} \right]$, the Jacobian matrix is at the fixed point $(\mathbf{x}^*,\mathbf{y}^*)$ is
\begin{equation*}
\left.
\begin{bmatrix}
\frac{\partial \mathbf{f}}{\partial \mathbf{x}} & \frac{\partial \mathbf{f}}{\partial \mathbf{y}} \\
\frac{\partial \mathbf{g}}{\partial \mathbf{x}} & \frac{\partial \mathbf{g}}{\partial \mathbf{y}}
\end{bmatrix}
\right|_{\mathbf{x}=\mathbf{x}^*,\mathbf{y}=\mathbf{y}^*}
\end{equation*}
The upper left submatrix is the same as the Jacobian of the subsystem with constant input from the rest of the system at the full system's fixed point, $d\mathbf{x}/dt = \mathbf{f}(\mathbf{x},\mathbf{y}^*)$ at $\mathbf{x} = \mathbf{x}^*$. It is generically different to that for the subsystem with zero input from the rest of the system, $d\mathbf{x}/dt = \mathbf{f}(\mathbf{x},0)$ at $\mathbf{x} = \mathbf{x}^*$.}

\subsection{Eigenvectors}
\label{sec:eigenvector}
The eigenvalues $\lambda_k$ of the Jacobian matrix, described in the previous section, represent dynamical modes of a social-ecological system. They are a system-level property and do not distinguish the contribution of individual state variables to the social-ecological system's dynamics. Each eigenvalue, however, has associated with it a pair of eigenvectors, a right eigenvector $\mathbf{v}^{(k)}$ and a left eigenvector $\mathbf{w}^{(k)}$ \citep{Wong1997} that do resolve state variables. We now describe how eigenvectors of the Jacobian matrix of a generalized model can be interpreted when analysing a social-ecological system.

The elements of a left eigenvector indicate the pattern of perturbations that must be applied to the state variables to excite exactly that mode. The elements of a right eigenvector indicate the extent to which the state variables are affected by that mode \citep{Kampmann2009,Aufderheide2013}. In a social-ecological system undergoing a B-tipping regime shift, the right eigenvector corresponding to the dominant eigenvalue can be used to predict the direction in which the system's state variables are expected to change at the time of the collapse (the "collapse direction").\footnote{In the case of an N-tipping regime shift, if the shocks on the system are known, the response of the system to the shocks could be predicted using a press perturbation approach \citep{Aufderheide2013}.}

Mathematically, the right eigenvectors of any unstable eigenvalues span the unstable eigenspace of the fixed point \citep{Solari1996}. The eigenspace, however, includes both positive and negative (parallel and antiparallel) multiples of the eigenvector; the dominant right eigenvector therefore can only indicate how state variables will move \emph{relative to each other}. Resolving absolute directions would require analyzing the model to higher orders than the linear analysis usually provided by generalized modeling \citep{Zumsande_2011}.

An additional caveat is that eigenvectors are a linear analysis, only predicting the initial direction of the collapse as the social-ecological system leaves its previous attractor. The right eigenvector is therefore unlikely to be a reliable estimator for the magnitude of the change in state variables, and should only be used to predict direction. Even the direction of change, however, could easily switch due to non-linear effects.

\citet{Aufderheide2013} also proposed indicators based on a weighted combination of the eigenvectors from different modes. They define the \emph{sensitivity} of state variable $i$ to the rest of the system as $\log\left(-\sum_k |v_i^{(k)}|/\lambda_k\right)$ and the \emph{influence} of state variable $i$ on the rest of the system as $\log\left(-\sum_k |w_i^{(k)}|/\lambda_k\right)$. In a model of a social-ecological system, sensitivity could possibly be used as a measure of how strongly the state variables would respond to an intervention, and influence for how strongly an intervention on that state variable would affect the rest of the system.
 
In the case of the Baltic cod fishery it is known that the biomass of sprat increased dramatically at the same time as the collapse of the cod stock. The known trajectories of sprat and other state variables in the cod fishery were compared against the predicted collapse direction as a validation test for the cod fishery model \citep{LadeRESULTS}. 

\subsection{Feedback loops}
\label{sec:feedback}
Regime shifts in social-ecological systems are often described in terms of a shift in the dominant feedback loops within the system \cite{Biggs_2012}. Feedback loops are pathways within a dynamical system in which changes in the value of any state variable in the loop affect the value of the same state variable, via the chain of variables indicated by the loop \citep{Sterman2000}. Feedback loops are an intuitive method for understanding the origin of a social-ecological system's dynamics. They are an appealing concept for understanding how system-level dynamics emerge from the interactions of specific processes in the social-ecological system.

`Reinforcing', `positive' or `unstable' loops magnify the initial disturbance, while `balancing', `negative' or `stable' loops reverse the effect of the disturbance and tend to return the system to equilibrium. Feedback loops of a `non-reduced' system include all intermediate variables in the loop description, while in `reduced' systems only the state variables are indicated \citep{Kampmann2012}.  

Feedback loops, however, are notoriously hard to systematically analyze \citep{Kampmann2009}. One of the most rigorous methods so far developed for analyzing feedback loops is Loop Eigenvalue Elasticity\footnote{The elasticities in LEEA and the elasticity type of generalized parameter share a common mathematical origin but their roles in model analysis and parameterization, respectively, should not be confused.} Analysis (LEEA) \citep{Kampmann2009,Kampmann2012}. The first step in applying LEEA to the Jacobian matrix produced by a generalized modeling analysis is identifying a complete set of feedback loops for the system. An immediate difficulty is that there is no unique `complete' set of feedback loops. The Shortest Independent Loop Set (SILS) algorithm of Oliva \citep{Oliva2004} produces the smallest set of loops that are `independent' of each other. An additional difficulty is that even within a particular loop set, the loops may not be very meaningful. For example, a pair of `phantom loops' may have large elasticities but always cancel each other out \citep{Kampmann2009}. 

The second step in LEEA is estimating the sensitivity of a chosen eigenvalue (the dominant eigenvalue is often the choice made) to the strength of each loop. The elasticity of the eigenvalue to individual links is first estimated using an analytical approximation \citep{Magnus1985,Kampmann2012}. The membership of links in the different loops is also known. The elasticity of the eigenvalue to loop strength can then be inferred \citep{Kampmann2012}.

In the Baltic Sea cod fishery case study, LEEA with SILS was applied to the (non-reduced) ecological and social subsystems as well as the social-ecological system \citep{LadeRESULTS}. Loop influence\footnote{Not to be confused with the influence based on the eigenvector analysis of \citet{Aufderheide2013} described above.}, rather than elasticity, was calculated as this avoids singular results near zero eigenvalue \citep{Kampmann2009}. Meaningful results were obtained in the ecological and social-ecological analysis, but the results for the social system were harder to interpret due to the model's structure: the feedback loops in the model's social system all overlap, with alternative pathways only in two, separate, sections. Focusing instead on the alternative pathways, and calculating sensitivity of the eigenvalue to individual links within those pathways, gave better insight into the dynamics of the social system than analyzing the entire feedback loop.

\subsection{Uncertainty analysis}
\label{sec:uncertainty}
A strength of generalized modeling as a tool for social-ecological system analysis is its ability to deal with uncertainty in the functional forms of processes in the social-ecological system. Analysis of the social-ecological system can therefore proceed without detailed knowledge of the functional forms, instead requiring only estimates of these generalized parameters. Rigorous model analysis, however, quantifies uncertainties in knowledge of the system (specifically, in the values of the generalized parameters) and assesses their effects on the model outputs. 

In the Baltic cod fishery model, uncertainties in the generalized parameters were estimated as described in Sec. \ref{sec:parameterization} above. The parameters' uncertainties, which were additionally assumed to be uncorrelated, were propagated through to the model outputs (such as eigenvalues) using Monte Carlo sampling. The assignation of parameter uncertainties, being possibly the first treatment of uncertainties in a generalized model, was deliberately chosen to be simplistic and transparent, and is likely to overestimate uncertainties. More sophisticated methods may yield more accurate uncertainty estimates. In particular, Bayesian approaches may be well suited to treat the mix of quantitative and qualitative information that is often available in a generalized modeling analysis.

\subsection{Sensitivity analysis}
\label{sec:sensitivity}
Sensitivity analyses examine how model outputs depend on individual model components, complementing analyses of the dependence of model output on larger model structures such as feedback loop analysis and system decomposition. Sensitivity analyses also complement uncertainty analysis in that they can apportion output uncertainty to different inputs \citep{Saltelli2008}.

Sensitivity analyses of complex models are often local due to computational cost, but local sensitivity analyses can be uninformative when the model output (here, stability) is a highly nonlinear function of its inputs (here, the generalized parameters). Global sensitivity analysis, where the entire parameter space is sampled, is generally preferable \citep{Saltelli2008}. The calculation of the Jacobian matrix in a generalized model, where it is immediately obtained from the generalized parameters, is much faster than in conventional simulation models, where the location of the fixed point must be re-evaluated for each combination of input parameters, and allows for a global sensitivity analysis even with large numbers of parameters.

In the Baltic Sea case study, variance-based sensitivity analysis \citep{Saltelli2008}, which is a type of global sensitivity analysis, was applied to the social-ecological system during the cod boom. Variance-based sensitivity analysis decomposes the variance in the model output into contributions from each input parameter. Two sensitivity measures are usually reported for each parameter: the `main effect index', the effect of varying that parameter alone, averaged over possible changes in the other parameters; and the `total effect index', the effect of that parameter together with the effects of all its interactions with other parameters \citep{Saltelli2008}. The analysis was computationally fast, even though the model had over 100 parameters. The variance-based sensitivity analysis was complemented with a conventional local derivative sensitivity analysis to obtain information on the direction in which a parameter affects the stability.

\subsection{Model experiments}
\label{sec:experiments}
The final type of generalized model analysis we introduce is model `experiments', in which the structure of the social-ecological model itself is modified, or where generalized parameters are adjusted to values outside their empirically determined range. The purposes of such experiments could include testing the structural assumptions that constitute the social-ecological model, or exploring model behavior in plausible but hypothetical alternative model structures.

Although generalized modeling allows additional interactions and feedback loops to be added very easily, care must be taken regarding assumptions on the fixed point of the social-ecological model during experiments. If a model experiment were to involve changes in the fixed point of the system, the values of generalized parameters in the model will generically also change, as they depend strongly on the location of the fixed point. In theoretical studies where a large range of generalized parameter values are considered, this is not a problem, as the theoretical range is likely to also apply to the new fixed point. In empirical studies, however, where generalized parameters are estimated at a specific fixed point (such as the cod boom), the previous estimate of the generalized parameter may not be accurate at the new fixed point.

We therefore recommend, in empirical studies, interpreting the model experiments as exploring a situation where \emph{the system remains at the same fixed point but the processes in the model are configured differently}. Unlike typical simulation models, the generalized model is not used to extrapolate system dynamics to hypothetical regions of state space for which there is no empirical data available, but rather to explore the effects of different process configurations at the same location in state space.

For example, in the Baltic cod fishery case study where model experiments were used to explore the effects of fishery regulations \citep{LadeRESULTS}, to ensure that the generalized parameters remain valid the experiments were interpreted as follows. The experiments explored an alternative Baltic Sea fishery, where all state variables (fish stocks, fisher effort levels, and fleet sizes) were the same as during the actual boom, but where regulation was an additional, strong factor influencing fisher decision making, implying compensatory changes in the strengths of other processes. A hypothetical situation where regulation had succeeded in reducing fishing effort and catch, moving the fishery into a new state, could not be investigated without requiring new estimates or unjustified extrapolations of generalized parameters.

\section{Discussion}
\subsection{Generalized modeling of empirical social-ecological systems}
Studies of social-ecological systems are frequently either in-depth case studies that can represent the complexity of a specific case \citep{Berkes__1998}, or theoretical studies that aim to represent general mechanisms and make general statements concerning social-ecological systems \citep{Biggs2009,Carpenter1999,Lade2013}. The results of case studies, however, can be difficult to generalize to other situations, and the theoretical studies can lack the complexity needed to be relevant to real-world policy-making situations. Modeling approaches at an intermediate level of complexity are required, which can take the complexity of contextual factors into account while also being able to provide system-level, generalizable explanations for the phenomena under study. Generalized modeling promises to be a method that provides such an approach. As the Baltic cod fishery example demonstrates, it can efficiently handle case-specific social and ecological complexity while also rigorously and rapidly analyzing the model at a system level.

The generalized modeling approach can function in the presence of limited data and a variety of quantitative and qualitative knowledge. Although they may help yield a more accurate model, long time series are not necessary to calibrate model functions in a conventional dynamical systems model, or to calculate stability via time series statistics in an early warning signal approach \citep{Dakos_PTRSB_2014}. Only point properties of the model's processes are required: the so-called generalized parameters that parameterize the system's Jacobian matrix, are easily interpreted and can often be extracted from qualitative knowledge about the social-ecological system.  At the same time, the lack of functional forms means that generalized models cannot simulate time series, an output that is usually expected of dynamical models. Instead generalized models should be applied to research questions involving qualitative dynamical patterns---such as the boom and collapse in the Baltic cod fishery, as demonstrated above. The qualitative dynamical patterns could be the final research question, as in the Baltic Sea case study, or they could a first modeling exercise preceding development of a more complicated model.

Integrating knowledge from traditionally isolated disciplines (such as the natural and social sciences) is another important challenge for social-ecological modeling. Causal loop diagrams are widely used, especially in collaborative or participatory contexts, to assess feedbacks within a social-ecological system \citep{Sendzimir2008,Downing2014}, but the `modeling' process often terminates there, due to the difficulty of developing a quantitative social-ecological model. As shown in the Baltic case study, generalized modeling is a promising tool for developing a causal loop diagram into a quantitative model in collaborative or participatory settings. 

A generalized model also can be efficiently subjected to several different forms of analysis, as shown here. A key output of a generalized modeling analysis is system stability (eigenvalues). Generalized modeling can also analyze the effects on stability of the social and ecological subsystems (system decomposition), individual feedback loops (feedback loop analysis) and individual processes or parameters (sensitivity analysis). Each of these analyses develop understanding of how social-ecological system dynamics emerge from individual processes. The effects of these dynamics on particular state variables can also be studied (eigenvector analysis). Several of these methods had not been applied to a generalized model before the Baltic case study. We specifically highlight feedback loop analysis as a promising tool, although there remain significant methodological challenges.

Differences between generalized models, conventional empirically-parameterized dynamical system models and theoretical models are summarized in Table \ref{tab1}. Key limitations of the generalized modeling approach, compared to direct numerical simulation of a dynamical system, are that: a generalized model cannot produce time series output, which may obstruct intuitive understanding of the model; lack of time series output may also obstruct empirical validation of the model; and that in the current formulation of the approach a fixed point is assumed. Limitations of the methods of analysis used here include: use of eigenvalues as a stability measure is susceptible to `localized modes'; eigenvectors can only indicate directions of change of state variables relative to each other, and are a linear analysis, only predicting the initial value of change; and in feedback loop analysis there is no unique method of defining a minimal loop set, and results can sometimes be difficult to interpret.

\begin{sidewaystable}
\caption{Comparison of generalized modeling with conventional empirical dynamical system models and theoretical dynamical system models\label{tab1}}
\small 

\begin{tabular}{p{3cm}p{3cm}p{3cm}p{3.5cm}p{3cm}}
Type of study &  Empirical structural information required & Empirical data required &  Most expensive computation for stability & Primary output \\
\hline 
Theoretical dynamical model & Not empirical & Not empirical & Numerical simulation of dynamical system & Stability or time series \\
Empirical dynamical model & State variables and processes & Functional forms and their parameters & Numerical simulation of dynamical system & Time series \\
Generalized model & State variables and processes & Generalized parameters & Eigenvalue comptuation & Stability
\end{tabular}

\end{sidewaystable}

\subsection{Lessons learned from Baltic model}
Here, we reflect on our experiences in using generalized modeling for the Baltic cod fishery social-ecological model \citep{LadeRESULTS}.

The construction of the conceptual model was a collaborative process involving researchers from several disciplines. During this phase we sought to concentrate almost exclusively on the social side of the system, as it had been less thoroughly studied than the ecological system. In contrast, when converting the causal loop diagram into a generalized model and estimating the generalized parameters, most difficulties were encountered in the ecological system, in particular the modeling of cod. Ultimately, appropriately modeling cod population dynamics required four state variables (adult and juvenile number and biomass) rather than one and the introduction of a model parameter that was the second-strongest contributor to uncertainty in the output \citep{LadeRESULTS}. Unlike the first phase of the model development process, where the lesser studied components (the social subsystem) attracted more attention, in this phase of the modeling process the better studied components (the ecological subsystem) attracted more attention. We speculate this is because constructing the model to be consistent with existing knowledge of the system and existing modeling work requires more effort for more well-known components. 

`Model experiments' on a generalized model were also introduced. We argued that model experiments are most easily performed such that the experiment does not change the location of the model's fixed point, but rather explores the effects of alternative configurations of processes that lead to the same fixed point. This can, however, restrict the usefulness of these `experiments'. In the Baltic fishery model it would have been useful to consider experiments where regulations had reduced fishing effort, and to test whether the boom would have been sustainable under reduced effort, for example. This is not possible in generalized modeling without re-estimating generalized parameters for the entire system at its new fixed point. On the other hand, it is advantageous that generalized modeling forces clarity on what model predictions are extrapolations and what are based on existing operating points. Extrapolations to new operating points could be made within a generalized modeling framework, given sufficient information to justify those extrapolations.

\subsection{Future developments of generalized modeling}
Finally, we speculate on important areas in which generalized modeling's development would facilitate its application to empirical social-ecological systems.

An important step of a generalized modeling analysis procedure is to assume the existence of a fixed point, so that the eigenvalues of the Jacobian matrix can be used as indicators of the stability of the fixed point. The Jacobian matrix, and objects computed from it such as local Lyapunov exponents \citep{Abarbanel1991} and finite-time Jacobian matrices \citep{Cvitanovic2012}, can however yield information about the qualitative dynamics of the system even when it is not near equilibrium. These quantities are generally used to estimate the rate of divergence of initially nearby trajectories. The quintessential situation is chaos where the rate of divergence is positive. We therefore speculate that a generalized modeling approach could be used to characterize the dynamics of a system even in the absence of a fixed point.

Generalized modeling analyses are fast to perform as only the eigenvalues of a matrix need to be computed, compared to solving a set of differential equations in a traditional simulation-based approach. Coding the generalized model was however also relatively time consuming, as the Jacobian needed to be derived manually (with some limited automation, as described above) and all analysis techniques coded manually. In order for generalized modeling to be widely used, a computer program with graphical interface similar to system dynamics' Vensim would be highly beneficial. The generalized modeling procedure in a theoretical context---construction of the generalized model, parameterization, and calculation of stability or bifurcations---is sufficiently mature to warrant such an effort. The present article shows, however, that methods important for applying generalized modeling to large-scale empirical systems, such as uncertainty analysis, sensitivity analysis and feedback loop analysis, still need to be refined.

\section{Conclusions}
Understanding the world's natural resource use and human wellbeing and development requires a specific type of systems thinking: social-ecological systems, in which natural and human systems are recognized as an interconnected and interdependent system \citep{Berkes__1998,Schlueter_NRM_2012}. We have elaborated the potential for generalized modeling, a type of dynamical systems modeling, to be used for empirical studies of social-ecological systems. A key advantage of generalized modeling is that the functional forms of the often poorly known social and ecological processes are not required. Even in the presence of this uncertainty generalized modeling provides formal mathematical tools with which to analyze the social-ecological model. Like other systems approaches, generalized modeling is also suitable for use in a collaborative or participatory mode. Although more development is required, generalized modeling is a promising tool for both researchers and stakeholders to rapidly gain an understanding of the resilience and qualitative dynamics of a social-ecological system.

\section*{Acknowledgements}
The research leading to these results has received funding from the European Research Council under the European Unions Seventh Framework Programme (FP/2007-2013)/ERC grant agreement no. 283950 SES-LINK, Project Grant 2014-589 from the Swedish Research Council Formas, and a core grant to the Stockholm Resilience Centre by Mistra. Maja Schl\"uter provided valuable comments on the manuscript.

\section*{References}
\bibliographystyle{apalike}
\bibliography{methods}

\end{document}